\documentclass[12pt,a4paper]{article}
\usepackage{amsfonts,latexsym}
\usepackage{amsmath,amssymb}
\usepackage{graphicx,color}
\usepackage{multirow}
\usepackage{slashbox}
\numberwithin{equation}{section} \oddsidemargin 0 mm \evensidemargin
0 mm \topmargin -10 mm \textheight 215 mm \textwidth 163 mm

\renewcommand{\thefootnote}{\fnsymbol{footnote}}
\newcommand{\nn}{\nonumber}

\begin{document}
\vspace{12mm}

\begin{center}
{{{\Large {\bf Critical gravity in the  Chern-Simons modified gravity}}}}\\[10mm]

{Taeyoon Moon$^{a}$\footnote{e-mail address: tymoon@sogang.ac.kr}
and  Yun Soo Myung$^{b}$\footnote{e-mail address:
ysmyung@inje.ac.kr},
}\\[8mm]

{{${}^{a}$ Center for Quantum Space-time, Sogang University, Seoul, 121-742, Korea\\[0pt]
${}^{b}$ Institute of Basic Sciences and School of Computer Aided Science, Inje University Gimhae 621-749, Korea}\\[0pt]
}
\end{center}
\vspace{2mm}

\begin{abstract}
We perform the perturbation analysis of  the Chern-Simons modified
gravity around  the AdS$_4$ spacetimes (its curvature radius $\ell$)
to obtain the critical gravity. In general, we could not obtain an
explicit form of perturbed Einstein equation which shows a massive
graviton propagation clearly,  but for the Kerr-Schild perturbation
and Chern-Simons coupling $\theta=kx/y$, we find the AdS wave as a
single massive solution to the perturbed Einstein equation. Its mass
squared is given by $M^2=[-9+(2\ell^2/k-1)^2]/4\ell^2$. At the
critical point of $M^2=0(k=\ell^2/2)$, the solution takes the
log-form and the linearized excitation energies vanish.
\end{abstract}
\vspace{5mm}

{\footnotesize ~~~~PACS numbers: }

\vspace{1.5cm}

\hspace{11.5cm}{Typeset Using \LaTeX}
\newpage
\renewcommand{\thefootnote}{\arabic{footnote}}
\setcounter{footnote}{0}


\section{Introduction}
The search for a consistent quantum gravity is mainly  being
suffered from obtaining a renormalizable and unitary quantum field
theory. Stelle has first introduced curvature squared terms of
$a(R_{\mu\nu}^2-R^2/3)+bR^2$ in addition to the Einstein-Hilbert
term of $R/2\kappa$~\cite{Stelle}. If $ab\not=0$, the
renormalizability was achieved, but the unitarity was violated
unless $a=0$. This clearly shows that the renormalizability and
unitarity exclude to each other. In other words, the
renormalizability requires  8 DOF (2 massless graviton, 5 massive
graviton from $a$-term, and 1 massive scalar from $b$-term), whereas
 the unitarity imposes 3 DOF (2 massless graviton and 1 massive
scalar). Although the $a$-term of providing massive graviton
improves the ultraviolet divergence, it induces ghost excitations
which jeopardize the unitarirty. In this sense, a first test for the
quantum gravity is to require the unitarity, which means that there
are no tachyon and ghost in its particle contents.

To this end, we would like to comment that  the critical gravities
as candidates for quantum gravity were recently investigated in the
AdS spacetimes~\cite{LP,DLL,AF,BHRT,PR,MKP,Myung}. At the critical
point, a degeneracy takes place and massive gravitons coincide with
either massless gravitons ($D>3$) or pure gauge modes ($D=3$).
Instead of massive gravitons, an equal amount of logarithmic modes
appears in the theory~\cite{BHML}: 1 DOF for topologically massive
gravity (TMG)~\cite{GJ}, 2 DOF for new massive
gravity~\cite{LS,GH,MKMP}, 5 DOF for higher curvature gravity in
4D~\cite{BHRT}. In general, we have $D(D+1)/2-(D+1)$ DOF for massive
graviton. However, the non-unitarity issue of the log-gravity is not
still resolved~\cite{LP,PR}, indicating that any log-gravity suffers
from the ghost problem. Furthermore, the critical gravity on the
Schwarzschild-AdS black hole has suffered from the ghost problem
when the cross term $E_{\rm cross}$ is non-vanishing~\cite{LLL}.

In this work, we introduce a Lorentz-violating theory of
Cherns-Simons modified gravity~\cite{jackiw}. A silent feature of
this theory is the presence of a constant vector $v_c$ which spoils
the isotropy of spacetime (CPT-symmetry) and is coupled to the
Pontryagin density of ${}^*RR$. Motivation of considering
Cherns-Simons modified gravity is  twofold in Minkowski spacetimes:
one is its close connection to the TMG  which accommodates a single
massive graviton in three dimensions~\cite{DJT} and the other is the
crucial dependence of massive graviton on a choice of constant
vector $v_c$. It was shown that a timelike vector of
$v_c=(\mu,\vec{0})$ did not provide any massive mode, leaving
massless graviton with 2 DOF, while a spacelike vector
$v_c=(0,\vec{v})$ yielded a massive graviton with 5
DOF~\cite{Boldo}. However, the authors~\cite{PereiraDias} have shown
that the only tachyon- and ghost-free model is the case of timelike
vector $v_c=(\mu,\vec{0})$, giving  2 DOF. This implies that the
role of Chern-Simons term is unclear to show its propagating DOF.

Here we wish to perform the perturbation analysis of  the
Chern-Simons modified gravity around  the AdS$_4$ spacetimes to
obtain the critical gravity, instead of Minkowski spactimes. Under
the transverse and traceless gauge, we could not obtain a compactly
third-order perturbed equation which shows a massive graviton with 5
DOF,  but for the Kerr-Schild perturbation with spacelike vector
$v_c=k(0,0,1/y,-x/y^2)$, we find the AdS wave as a single massive
graviton   propagating on AdS$_4$ spacetimes. This was found as a
solution to the Einstein equation~\cite{beato}. This (1 DOF)
contrasts to propagating DOF  of graviton in Minkowski spacetimes.
At the critical point of $k=\ell^2/2$, the solutions takes the
log-form and the linearized excitation energies vanish, which
indicates a feature of critical gravity.

\section{Chern-Simons modified gravity}
Let us first consider the Chern-Simons modified gravity in four
dimensions with a cosmological constant ($\Lambda$) whose action is
given by
\begin{eqnarray}
S=\frac{1}{16\pi G}\int d^4 x\sqrt{-g}
\Big\{R-2\Lambda+\frac{\theta}{4}{}^{*}RR \Big\}\label{Action}
\end{eqnarray}
where $\theta$  \footnote{$\theta$ is a diffeomorphism breaking
parameter and it will be   fixed  by the equation of motion.
Therefore, it is hard to be considered as  a Lagrange multiplier. In
the Chern-Simons modified Maxwell theory, $\theta$ can be fixed as
$\mu t$ which yields the modified Ampere's law~\cite{jackiw}. At
this stage, one may ask the question  ``can we call any model
without diffeomorphism as gravity?''. In order to answer it, we
remind  the feature of  the gravitational Chern-Simons modified
theory~\cite{jackiw}. Here the diffeomorphism breaking  is being
realized from the fact that the covariant divergence of the
four-dimensional Cotton tensor is non-zero [see Eq.(\ref{bian})], in
contrast to the case  in three dimensions.  Therefore, a consistency
condition on this theory is that ${}^{*}RR=0$  for
$\nabla_{b}\theta\neq0$ (the theory reduces to the general
relativity for $\nabla_{b}\theta=0$ because of $C_{ab}=0$). In this
sense, diffeomorphism symmetry breaking is suppressed dynamically
for the case of ${}^{*}RR=0$ (e.g., Schwarzschild black hole or
AdS$_4$ spacetimes), even if it may occur at the action level.} is
an external function of spacetime and
${}^{*}RR={}^{*}R^{a~cd}_{~b}R^{b}_{~acd}$ is the Pontryagin density
with
\begin{eqnarray}
{}^{*}R^{a~cd}_{~b}=\frac{1}{2}\epsilon^{cdef}R^{a}_{~bef}.
\end{eqnarray}
In this expression, $\epsilon^{cdef}$ denotes the four-dimensional
Levi-Civita tensor. Varying for $g_{ab}$ on the action
(\ref{Action}) leads to
 the Einstein equation which takes the form
\begin{eqnarray} \label{equa1}
R_{ab}-\frac{1}{2}g_{ab}R+\Lambda g_{ab}+C_{ab}=0
\end{eqnarray}
where $C_{ab}$ is the four-dimensional Cotton tensor given by
\begin{eqnarray}\label{cotton}
C_{ab}=\nabla_{c}~\theta~\epsilon^{cde}_{~~(a}\nabla_{|e|}R_{b)d}+\frac{1}{2}\nabla_c\nabla_d
~\theta~\epsilon_{(b}^{~~cef}R^{d}_{~~a)ef}.
\end{eqnarray}
 Note that $C_{ab}$ is a traceless and symmetric tensor. As a result of applying Bianchi identity to (\ref{equa1}), one has
 \begin{equation} \label{bian}
 \nabla^aC_{ab}=\Big[\frac{\nabla_b\theta}{8}\Big]~ {}^*R_{acdf} R^{acdf}.
 \end{equation}
On the other hand, one finds  that Eq.(\ref{equa1}) has an AdS$_4$
solution in which
 the Riemann tensor, Ricci tensor and Ricci scalar of the AdS$_4$ spacetimes are given by
\begin{eqnarray}
\bar{R}_{abcd}=\frac{\Lambda}{3}(\bar{g}_{ac}\bar{g}_{bd}-\bar{g}_{ad}\bar{g}_{bc}),~~~\bar{R}_{ab}=\Lambda
\bar{g}_{ab},~~~\bar{R}=4\Lambda.
\end{eqnarray}
Here ``overbar'' denotes the background AdS$_4$-metric
$\bar{g}_{ab}$.

In order to obtain perturbation equations, we introduce the
perturbation around the the background metric as
\begin{eqnarray} \label{m-p}
g_{ab}=\bar{g}_{ab}+h_{ab}.
\end{eqnarray}
The linearized equation to (\ref{equa1}) can be written by
\begin{eqnarray}\label{pert0}
\delta R_{ab}(h)-\frac{1}{2}g_{ab}\delta R(h)-\Lambda h_{ab}+\delta
C_{ab}(h)=0,
\end{eqnarray}
where the linearized tensor $\delta R_{ab}(h),~\delta R(h),$ and
$\delta C_{ab}(h)$ take the form
\begin{eqnarray}\label{cottonp0}
\delta R_{ab}(h)&=&\frac{1}{2}\left(\bar{\nabla}^{c}\bar{\nabla}_a
h_{bc}+\bar{\nabla}^{c}\bar{\nabla}_b
h_{ac}-\bar{\nabla}^2h_{ab}-\bar{\nabla}_a \bar{\nabla}_b h\right)\nn\\
\delta
R(h)&=&\bar{\nabla}^{a}\bar{\nabla}^{b}h_{ab}-\bar{\nabla}^2h-\Lambda
h\nn\\
\delta
C_{ab}(h)&=&\Bigg[\frac{1}{2}v_{c}~\epsilon^{cde}_{~~~a}\left(\bar{\nabla}_{e}\delta
R_{bd}-\Lambda \bar{\nabla}_{e}h_{bd}\right)+\frac{1}{8}v_{cd}~
\epsilon_{b}^{~cef}\Big(\bar{\nabla}_e\bar{\nabla}_f
h^{d}_{~a}+\bar{\nabla}_e\bar{\nabla}_a h^{d}_{~f}\nn\\
&&-\bar{\nabla}_e\bar{\nabla}^d h_{af}-\bar{\nabla}_f\bar{\nabla}_e
h^{d}_{~a}-\bar{\nabla}_f\bar{\nabla}_a
h^{d}_{~e}+\bar{\nabla}_f\bar{\nabla}^d
h_{ae}\Big)\Bigg]+\Bigg[a\leftrightarrow b\Bigg]
\end{eqnarray}
with $v_c=\bar{\nabla}_c \theta$ and
$v_{cd}=\bar{\nabla}_c\bar{\nabla}_d \theta$. Imposing the
transverse and traceless (TT) gauge condition as
\begin{eqnarray}\label{gauge}
\bar{\nabla}_{a}h^{ab}=0,~~h=\bar{g}^{ab}h_{ab}=0
\end{eqnarray}
which takes into account the diffeomorphism~\cite{Tekin}
\begin{equation}
\delta_\xi h_{ab}=\bar{\nabla}_a\xi_b+\bar{\nabla}_b\xi_a,
\end{equation}
 the perturbation equation (\ref{pert0}) takes a simpler form
\begin{eqnarray}\label{pert}
-\frac{1}{2}\bar{\nabla}^2h_{ab}+\frac{\Lambda}{3}h_{ab}+\delta
C_{ab}=0.
\end{eqnarray}
Here the linearized tensor $\delta C_{ab}(h)$  is given by
\begin{eqnarray}\label{cottonp}
\delta
C_{ab}(h)&=&\Big[-\frac{1}{4}v_{c}~\epsilon^{cde}_{~~~a}\bar{\nabla}_e\bar{\nabla}^2h_{bd}+
\frac{\Lambda}{6}v_c~ \epsilon^{cde}_{~~~a}\bar{\nabla}_e
h_{bd}+\frac{1}{4}v_{cd}~\epsilon_{b}^{~cef}\Big(\bar{\nabla}_e\bar{\nabla}_a
h^{d}_{~f}-\bar{\nabla}_e\bar{\nabla}^d
h_{af}\Big)\Big]\nn\\&&\hspace*{0em}+\Big[a\leftrightarrow b\Big].
\end{eqnarray}
We observe that $\delta C_{ab}(h)$ takes still a complicated form,
depending $v_c$ and $v_{cd}$.

\section{AdS wave as perturbation}

It is important to note that the perturbation equation (\ref{pert})
has the dependency of $\theta$. For a choice of $\theta=t/\mu$
\cite{jackiw}, the  Cotton tensor (\ref{cotton}) reduces to the TMG
when choosing the Schwarzschild coordinates. However, in the AdS$_4$
spacetimes,  such a choice is not guaranteed  since the second term
$v_{ab}$  survives.  In the AdS$_4$ spacetimes, there  exists a
particular choice of $\theta$ \cite{beato} which makes $v_{cd}$
vanish. This choice of $\theta$ could be made   by choosing the
Poincare coordinates $(u,v,x,y)$ for the AdS$_4$ spacetimes:
\begin{eqnarray}\label{ansat}
\theta=k\frac{x}{y},~~~
\bar{g}_{ab}=\phi^{-2}\eta_{ab}=\frac{\ell^2}{y^2}\eta_{ab},
\end{eqnarray}
where $k$($>0$)  has the dimension of [mass]$^{-2}$, $\ell$ is the
AdS$_4$ curvature radius ($\ell^2=-3/\Lambda)$ and $\eta_{ab}$ is
\begin{eqnarray}
\eta_{ab}dx^{a}dx^{b}=2dudv+dx^2+dy^2.
\end{eqnarray}
Considering Eq.(\ref{ansat}),    Eq.(\ref{pert}) becomes
\begin{eqnarray}\label{Master1}
\Big(\bar{\nabla}^2-\frac{2}{3}\Lambda\Big)
\Big(h_{ab}+v_c~\epsilon^{cde}_{~~~(a}\bar{\nabla}_{|e|}h_{b)d}\Big)=0.
\end{eqnarray}
Alternatively, it leads to
\begin{eqnarray}\label{Master1-1}
\Big(\delta_{(a}^{a^{'}}\delta_{b)}^{d}
+\delta_{(a}^{a^{'}}v_{|c|}~\epsilon^{cde}_{~~~b)}\bar{\nabla}_e\Big)
\Big(\bar{\nabla}^2-\frac{2}{3}\Lambda\Big)h_{a^{'}d}=0
\end{eqnarray}
which is found  by using commutation  between two operations in
Eq.(\ref{Master1}). Here,  $v_c$ is given by
\begin{equation}\label{vc}
v_c=k\Bigg(0,0,\frac{1}{y},-\frac{x}{y^2}\Bigg) \end{equation} which
may generate the mass. In this case, $v_c$ is not a constant vector
but a vector field. We wish to  comment that Eq.(\ref{Master1}) is
an extended version in four dimensions when comparing with the TMG
\cite{strom}. In three dimensions, one  analyzes the perturbation
equation by using $D$-operator
\begin{equation}
\Big(D^{\mu/\tilde{\mu}}\Big)_{\alpha}^{~\beta}=\delta_{\alpha}^{~\beta}\pm
\frac{1}{\mu}\epsilon_{\alpha}^{~\gamma\beta}\bar{\nabla}_{\gamma}.
\end{equation}
However, it is not easy   to apply $D$-operator directly to Eq.
(\ref{Master1}) because $v_c$ is not a constant vector. In order to
see this case explicitly, we introduce
$\hat{D}^{\tilde{M}}$-operator in the AdS$_4$ spacetimes
\begin{equation} \label{dmoper}
\Big(\hat{D}^{M/\tilde{M}}\Big)^{ff'}_{ad}=\delta_{(a'}^{f}\delta_{d)}^{f'}
\pm\delta_{(a'}^{f}v_{|c|}~\epsilon^{cf'e}_{~~~~d)}\bar{\nabla}_e.
\end{equation}
Then, Eq.(\ref{Master1-1}) can be rewritten as
\begin{equation}
\Big(\hat{D}^{M}\Big)^{a'd}_{ab}\Big(\bar{\nabla}^2-\frac{2}{3}\Lambda\Big)h_{a'd}=0.
\end{equation}
Now we  use $\hat{D}^{\tilde{M}}\hat{D}^M$-operation
 to find
\begin{eqnarray}\label{DD}
&&\Big(\delta_{(a'}^{f}\delta_{d)}^{f'}
-\delta_{(a'}^{f}v_{|c'|}~\epsilon^{c'f'e'}_{~~~~~d)}\bar{\nabla}_e'\Big)
\Big(\delta_{(a}^{a'}\delta_{b)}^{d}
+\delta_{(a}^{a'}v_{|c|}~\epsilon^{cde}_{~~~~b)}\bar{\nabla}_e\Big)
h_{ff'}
\nn\\
&&\nn\\
&&\hspace*{-1.5em}=-4v^2\Big(\bar{\nabla}^2-\frac{2}{3}\Lambda
-\frac{1}{v^2}\Big)h_{ab}
+4v^{e'}v^{e}\bar{\nabla}_{e'}\bar{\nabla}_{e}h_{ab} -2\Lambda\theta
v^{e}\bar{\nabla}_{e}h_{ab}-2v^{e}v^{e'}\bar{\nabla}^2h_{ee'}g_{ab}
\nn\\
&&+\frac{8}{3}\Lambda v^{e}v^{e'}h_{ee'}g_{ab}+
\Big[-2v^{e'}v^{e}\bar{\nabla}_a\bar{\nabla}_{e}h_{e'b}+3v^{e}v_{a}\bar{\nabla}^2h_{eb}
-\frac{\Lambda}{3}\theta v^{e}\bar{\nabla}_{a}h_{eb}
+v^{e'}v^{e}\bar{\nabla}_a\bar{\nabla}_{b}h_{ee'}
\nn\\
&&-v^{e}v^{e'}\bar{\nabla}_{e'}\bar{\nabla}_{a}h_{eb}
-\frac{8}{3}\Lambda v^{e}v_{b}h_{ea}+(a\leftrightarrow b)\Big] =0
\end{eqnarray}
with $v^2\equiv v_ev^e$. In obtaining this, we have used the gauge
condition (\ref{gauge}).  At this stage,  it is  very difficult to
derive the massive second-order equation\footnote{ Assuming that all
terms except the first term of
$-4v^2\Big(\bar{\nabla}^2-\frac{2}{3}\Lambda
-\frac{1}{v^2}\Big)h_{ab}$ vanish, it has still a problem to derive
Eq.(\ref{massiv}).  This is because $v^2$ is not a constant scalar
as $\frac{k^2}{\ell^2}=\frac{1}{\mu^2}$  in the TMG,  but it is a
scalar function given by $v^2=\frac{k^2}{\ell^2}\frac{x^2+y^2}{
y^2}$.},
\begin{equation}\label{massiv}
\Big(\bar{\nabla}^2-\frac{2}{3}\Lambda-M^2\Big)h_{ab}=0\end{equation}
unless we choose  a simple form of the metric perturbation $h_{ab}$.

 In order to analyze Eq.(\ref{Master1}), we consider the AdS$_4$
wave  as the Kerr-Schild form
\begin{eqnarray}\label{metric1}
h_{ab}=2\varphi\lambda_{a}\lambda_{b},
\end{eqnarray}
where $\lambda_{a}$ is a null and geodesic vector whose form is
given by $\lambda_{a}=(1,0,0,0)$ and $\varphi$ is an arbitrary
function of coordinates $(u,v,x,y)$. To maintain the TT gauge
condition (\ref{gauge}),  one  confines  $\varphi$ to
$\varphi(u,x,y)$ by requiring the condition of
$\lambda_{a}\bar{\nabla}^{a}\varphi=0$ ($\to\partial_{v}\varphi=0$).
Plugging $h_{ab}=2\varphi\lambda_a\lambda_b$ into Eq.(\ref{Master1})
leads to
\begin{eqnarray} \label{adsteneq}
\lambda_{a'}\lambda_{d}\Big[\delta_{(a}^{a^{'}}\delta_{b)}^{d}
+\delta_{(a}^{a^{'}}v_{|c|}~\epsilon^{cde}_{~~~b)}
\Big(\frac{\partial_e\phi}{\phi}+\bar{\nabla}_e\Big)\Big]
\Big[\bar{\nabla}^2+\frac{2}{3}\Lambda+\frac{4}{\phi}\partial^{f}\phi~\partial_{f}\Big]\varphi=0.
\end{eqnarray}
At this stage, we introduce the separation of variables by
considering
\begin{equation}
\varphi(u,x,y)=U(u)X(x)Y(y).
\end{equation}
Taking into account $\lambda_a,~v_c,$ and $\phi$,
Eq.(\ref{adsteneq}) can be reduced to
\begin{eqnarray}\label{master3}
\Big[y\partial_y+x\partial_x+1-\frac{\ell^2}{k}\Big]
\Big[y^2(\partial_y^2+\partial_x^2)+2y\partial_y-2\Big]XY=0.
\end{eqnarray}
Note that the right bracket in Eq.(\ref{master3}) represents the
perturbation equation of the massless scalar which  corresponds to
the right parenthesis of massless tensor  in Eq.(\ref{Master1-1}).
On the other hand,  we expect that the left bracket in
Eq.(\ref{master3}) is related to the massive-mode equation as was
suggested  in three dimensions~\cite{strom}.  In order to obtain the
massive-mode (scalar) equation  from the left bracket in
Eq.(\ref{master3}),  we
 introduce  an  operator of  $y\partial_{y}+A$ with  $A$ an arbitrary
 constant. Furthermore, we  assume that $X(x)$=constant.
Then, we check that  the quadratic perturbation equation yields
\begin{eqnarray}
&&(y\partial_{y}+A)(y\partial_{y}+1-\ell^2/k)Y=0\nn\\
&&\to\Big[y^2\partial_y^2+y(2-\ell^2/k+A)\partial_y+A(1-\ell^2/k)\Big]Y=0,\label{eq1}
\end{eqnarray}
while the perturbation equation of the massive mode may take the
form
\begin{eqnarray}
&&\Big[\bar{\nabla}^2+\frac{2}{3}\Lambda
+\frac{4}{\phi}\partial^{a}\phi\partial_{a}-M^2\Big]\varphi=0\nn\\
&& \to \frac{1}{\ell^2}
\Big[y^2\partial_y^2+2y\partial_y-2-M^2\ell^2\Big]Y=0\label{eq2}.
\end{eqnarray}
Comparing Eq.(\ref{eq1}) with Eq.(\ref{eq2}),  we
find\footnote{There also exists the solution of
$(1-\sqrt{9+4\ell^2M^2})/2$. However, it violates the allowed region
of $M^2$,  $M^2<M_{BF}^2$ for $k>0$. This induces the tachyon
instability. Hence, we ignore this solution for the Chern-Simons
coupling $k>0$.}
\begin{eqnarray}\label{rel}
A=\frac{\ell^2}{k},~~\frac{\ell^2}{k}=\frac{1}{2}\Big(1+\sqrt{9+4\ell^2
M^2}\Big).
\end{eqnarray}
It is worth noting  that for real  $\ell^2/k$,  the allowed region
of $M^2$ is  given by
\begin{eqnarray}\label{mass}
M^2=\frac{1}{4\ell^2}\Big[-9+\left(\frac{2\ell^2}{k}-1\right)^2\Big]\ge
M^2_{\rm BF}=-\frac{9}{4\ell^2},
\end{eqnarray}
where $M^2_{\rm BF}$ corresponds to the Breitenlohner-Freedman (BF)
bound for a massive scalar  in AdS$_4$ spacetimes~\cite{BF}. This
occurs also for $k=2\ell^2$.  Importantly, in the critical limit of
$M^2\to0$, we obtain $k=\ell^2/2$ from Eq.(\ref{rel}). In addition,
we note that for $k>\ell^2/2$, we have an allowed bound for negative
$M^2$ (see Fig.1)
\begin{equation}
M^2_{\rm BF} \le M^2 < 0
\end{equation}
which was also derived from the tensor analysis in the higher
curvature gravity including the conformal gravity~\cite{LPP}.

\begin{figure*}[t!]
   \centering
   \includegraphics{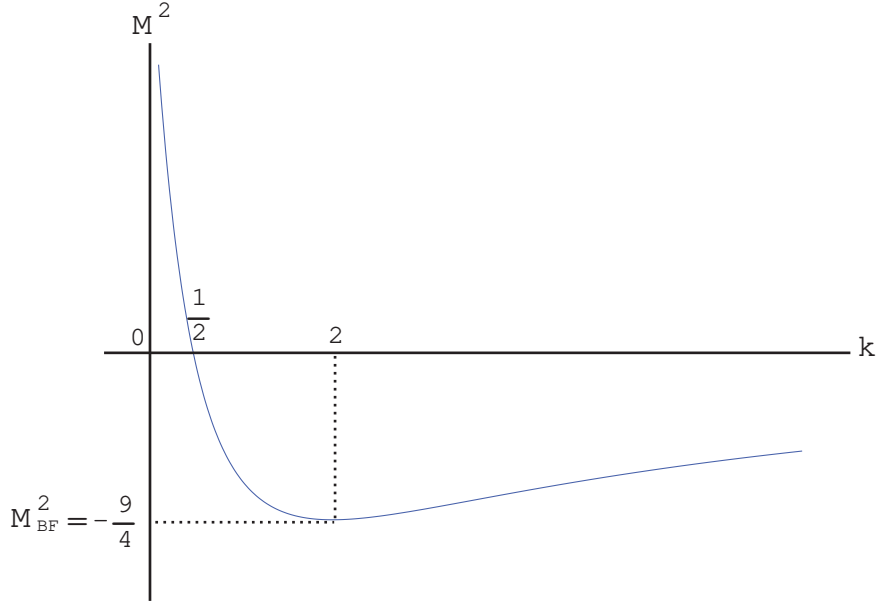}
\caption{$M^2$ graph as function of $k$ with  $\ell^2=1$.  For
$k>1/2$ ($M^2<0$), the AdS wave is a stable solution because it
satisfies the BF bound,  $M^2\ge M_{BF}^2=-9/4$. Hence we have the
stable region for  positive $k$ ($k>0$). The figure shows that the
critical point is located at $k=1/2$ which corresponds to $M^2=0$.
Also, in the limit of $k\to \infty$, it approaches $M^2=-2$.}
\label{CS-2}
\end{figure*}

Consequently, Eq.(\ref{master3}) reduces to  the third-order
equation for $y$
\begin{eqnarray}\label{master}
\Bigg[y\partial_y+\frac{1}{2}\Bigg(1-2\ell\sqrt{M^2+\frac{9}{4\ell^2}}~\Bigg)
\Bigg]\Big[y\partial_y-1\Big]\Big[y\partial_y+2\Big]Y=0.
\end{eqnarray}
Now we solve Eq.(\ref{master}) for two cases:\\
\\
(i) $k\neq\ell^2/2$  $(M^2\neq0)$
\begin{eqnarray}
\varphi(u,y)=U(u)Y(y)=c_1(u)y^{\frac{1}{2}[-1+2\ell\sqrt{M^2+9/4\ell^2}]}+c_2(u)\frac{1}{y^2}+c_3(u)
y,
\end{eqnarray}
which is a single massive solution in AdS$_4$ spacetimes.
\\
\\
(ii) $k=\ell^2/2$  ($M^2=0$)\\
In this case, Eq.(\ref{master}) degenerates as
\begin{eqnarray}\label{master2}
(y\partial_y+2)(y\partial_y-1)^2Y=0.
\end{eqnarray}
We obtain the solution as
\begin{eqnarray}\label{sol3}
\varphi(u,y)=U(u)Y(y)=c_4(u)y\ln(y)+c_5(u)\frac{1}{y^2}+c_6(u) y.
\end{eqnarray}
In this approach, $c_i(u)$ as functions of $u$ remain undetermined.

We note that the solution (\ref{sol3}) will be a half of the
solution obtained from  higher curvature gravity which gives  the
fourth-order perturbation equation at the critical
point~\cite{tekin}.  To see this more closely, we construct the
fourth-order equation instead of the third-order equation
(\ref{master})  by considering
\begin{eqnarray}\label{fourth}
\Big[(y\partial_y+\ell^2/k)(y\partial_y+1-\ell^2/k)\Big]
\Big[(y\partial_y-1)(y\partial_y+2)\Big]Y=0.
\end{eqnarray}
For $k\neq\ell^2/2$, the solution to Eq.(\ref{fourth}) is given by
$Y=Y_1+Y_2$ where $Y_1$ and $Y_2$ satisfy the following second-order
equations, respectively:
\begin{eqnarray}
&&\Big[(y\partial_y+\ell^2/k)(y\partial_y+1-\ell^2/k)\Big]Y_1=0,~~~
\Big[(y\partial_y-1)(y\partial_y+2)\Big]Y_2=0.
\end{eqnarray}
The corresponding solutions and combined solution  are
\begin{eqnarray}
&&Y_1(y)=d_1y^{\frac{1}{2}[-1+2\ell\sqrt{M^2+9/4\ell^2}]}+d_2
y^{-\frac{1}{2}[1+2\ell\sqrt{M^2+9/4\ell^2}]},\\
&&Y_2(y)=d_3y^{-2}+d_4 y,\\
&&\hspace*{-3em}\rightarrow
Y(=Y_1+Y_2)=d_1y^{\frac{1}{2}[-1+2\ell\sqrt{M^2+9/4\ell^2}]}+d_2
y^{-\frac{1}{2}[1+2\ell\sqrt{M^2+9/4\ell^2}]}+ d_3y^{-2}+d_4 y,
\end{eqnarray}
where $M^2$ appeared in (\ref{mass}) and $d_{i}$ are undetermined
constants. We note that although the solution form is the same as
found in the higher curvature gravity~\cite{tekin}, the mass squared
$M^2$ in (\ref{mass}) is different from that [(8) in~\cite{tekin}]
in the higher curvature gravity. At the critical point of $M^2=0$
($k=\ell^2/2$), the fourth-order equation reduces to
\begin{eqnarray}
\Big[\Big(y\partial_y-1\Big)\Big(y\partial_y+2\Big)\Big]^2Y=0,
\end{eqnarray}
whose solution is given by
\begin{eqnarray}
Y(y)=d_5y\ln(y)+d_6\frac{1}{y^2}+d_7 y+\frac{d_8}{y^2}\ln(y)
\end{eqnarray}
which shows that the last term is absent in (\ref{sol3}). This
solution is exactly the same found in the higher curvature
gravity~\cite{tekin}.

\section{Linear excitation energy}
In the perturbation analysis,  it is important  to check  whether
the ghost mode exists or not.  For this purpose, we  construct the
Hamiltonian of the action. Firstly, the quadratic action of $h_{ab}$
takes the form
\begin{eqnarray}\label{2action}
S^{(2)}&=&-\frac{1}{16\pi G}\int
d^4x\sqrt{-g}h^{ab}\Big[\delta\Big(R_{ab}-\frac{1}{2}g_{ab}R+\Lambda
g_{ab}\Big)+\delta C_{ab}\Big]\nn\\
&=&-\frac{1}{16\pi G}\int
d^4x\sqrt{-g}\Big[\frac{1}{2}(\bar{\nabla}^{c}h^{ab})(\bar{\nabla}_{c}h_{ab})
+\frac{\Lambda}{3}h^{ab}h_{ab}+\frac{1}{2}\epsilon^{cde}_{~~~a}
\Big(v_{ce}h^{ab}\bar{\nabla}^2h_{bd}\nn\\
&&+v_c\bar{\nabla}_eh^{ab}\bar{\nabla}^2h_{bd} +\frac{2}{3}\Lambda
v_c
h^{ab}\bar{\nabla}_eh_{bd}\Big)-\frac{1}{2}\epsilon_b^{~cef}\Big(v_{cde}
h^{ab}\bar{\nabla}_{a}h^{d}_{~f}+v_{cd}\bar{\nabla}_{e}h^{ab}\bar{\nabla}_{a}h^{d}_{~f}
\nn\\
&&-v_{cde}h^{ab}\bar{\nabla}^{d}h_{af}-v_{cd}\bar{\nabla}_{e}h^{ab}\bar{\nabla}^{d}h_{af}\Big)\Big].
\end{eqnarray}
From the action (\ref{2action}),  we define the conjugate momentum
given by
\begin{eqnarray}
&&\Pi^{ab}_{(1)}=-\bar{\nabla}^2h^{ab}-\frac{1}{2}\epsilon^{cd0a}v_c
\bar{\nabla}^2h^{b}_{~d}-\frac{\Lambda}{3}v_c\epsilon^{ca0}_{~~~d}h^{db}
+\frac{1}{2}\epsilon^{bc0f}v_{cd}\bar{\nabla}^{a}h^{d}_{~f}+\frac{1}{2}\epsilon_{f}^{~ceb}v_c^{a}
\bar{\nabla}_e h^{0f}\nn\\
&&\hspace*{1em}-\frac{1}{2}\epsilon_f^{~ceb}v_{c~e}^{~0}h^{af}
-\frac{1}{2}\epsilon^{bc0f}v_{cd}\bar{\nabla}^{d}h^{a}_{~f}-\frac{1}{2}\epsilon_f^{~ceb}v_c^{0}
\bar{\nabla}_{e}h^{af}+\frac{1}{2}\bar{\nabla}^{0}(\epsilon^{cae}_{~~~d}v_c\bar{\nabla}_{e}h^{db}\bar{g}^{00}).
\end{eqnarray}
Using the method of Ostrogradsky, we  find  the conjugate momentum
for the second-time derivative as
\begin{eqnarray}
\Pi^{ab}
_{(2)}=-\frac{1}{2}\bar{\nabla}^{0}(\epsilon^{cae}_{~~~d}v_c\bar{\nabla}_{e}h^{db}\bar{g}^{00}).
\end{eqnarray}
Then the Hamiltonian can be written by
\begin{eqnarray}\label{hamil}
H=\int
d^{4}x\Big(\dot{h}_{ab}\Pi^{ab}_{(1)}+\dot{K}_{ai}\Pi^{ai}_{(2)}\Big)-S^{(2)}
\end{eqnarray}
with $K_{ai}=\bar{\nabla}_{0}h_{ai}$.  Considering (\ref{ansat}) and
 (\ref{metric1}),  one finds  that the Hamiltonian (\ref{hamil})
is identically zero ($H=0$), irrespective of any solution form
$\varphi$. This means that there is no ghost for AdS waves.

 \vspace{1cm}

\section{Discussions}
In the Minkowski spacetimes, the ghost- and tachyon-free mode of
Chern-Simons modified gravity is just a massless graviton with 2
DOF~\cite{PereiraDias}. This amounts to the choice of a timelike
vector $v_c=(\mu,\vec{0})$.

In general, it is a formidable task to find a massive graviton with
5 DOF in the AdS$_4$ spacetimes because its linearized equation is a
very complicated form, compared to the TMG, showing  a single
massive scalar~\cite{strom}.  However, choosing $v_c$ as a vector
field (\ref{vc}) which makes the perturbation equation simple and
then, the Kerr-Schild perturbation (\ref{metric1}),  we have a
single massive scalar $\varphi$ propagating on the AdS$_4$
spacetimes. This is ghost-free and tachyon-free if the mass squared
(\ref{mass}) satisfies the BF bound $M^2 \ge M^2_{\rm BF}$. Even for
the negative bound of $-M^2_{\rm BF} \le M^2 < 0$, there is no
tachyon instability (no exponentially growing modes)~\cite{LPP}. At
the critical point of $M^2=0(k=\ell^2/2$), we have found the
log-form without ghost, which is the half solution found in the
higher curvature gravity~\cite{tekin}.

However, it seems difficult to derive a massive graviton with 5 DOF
propagating in the AdS$_4$ spacetimes from the Chern-Simons modified
gravity, compared to the higher curvature gravity~\cite{BHML}. This
is mainly because  massive excitations depend  critically on  the
choice of coupling field $v_c (\theta)$.

 \vspace{1cm}

{\bf Acknowledgments}

This work was supported by the National Research Foundation of Korea
(NRF) grant funded by the Korea government (MEST) through the Center
for Quantum Spacetime (CQUeST) of Sogang University with grant
number 2005-0049409. Y. Myung  was partly  supported by the National
Research Foundation of Korea (NRF) grant funded by the Korea
government (MEST) (No.2011-0027293).

\end{document}